\documentclass[12pt]{article}
\usepackage{color}
\usepackage{graphicx}
 \usepackage{amsmath,amssymb,array,calc}%rotating,
\usepackage{bm}

\definecolor{darkgreen}{rgb}{0,0.55,0}

\newcommand{\bea}{\begin{eqnarray}}
\newcommand{\eea}{\end{eqnarray}}
\newcommand{\be}{\begin{equation}}
\newcommand{\ee}{\end{equation}}
\newcommand{\nn}{\nonumber}

\def\s{\sigma}

\def\revise#1       {\raisebox{-0em}{\rule{3pt}{1em}}%
                     \marginpar{\raisebox{.5em}{\vrule width3pt\
                     \vrule width0pt height 0pt depth0.5em
                     \hbox to 0cm{\hspace{0cm}{%
                     \parbox[t]{4em}{\raggedright\footnotesize{#1}}}\hss}}}}

\def\sqr#1#2{{\vcenter{\vbox{\hrule height.#2pt
 \hbox{\vrule width.#2pt height#1pt \kern#1pt
 \vrule width.#2pt}\hrule height.#2pt}}}}

\catcode`\@=12

\begin{document}

%%%%% number equations by section %%%%%%%%
\makeatletter \@addtoreset{equation}{section} \makeatother
\renewcommand{\theequation}{\thesection.\arabic{equation}}

\renewcommand\baselinestretch{1.25}
\setlength{\paperheight}{10in}
\setlength{\paperwidth}{9.5in}

\begin{titlepage}

%\version\versionno

%

\vskip 4cm

 \begin{center}
{\bf \large Constraints and Generalized Gauge Transformations
on Tree-Level Gluon and Graviton Amplitudes}
\end{center}
\vskip 1cm

\centerline{\large  Diana Vaman\textsuperscript{*}, York-Peng Yao\textsuperscript{$\dagger$} {}\footnote{E-mail addresses: dv3h@virginia.edu, yyao@umich.edu}
}

\vskip .5cm
\centerline{\it \textsuperscript{*} Department of Physics, The University of Virginia}
\centerline{\it Charlottesville, VA 22904, USA}
\centerline{\it \textsuperscript{$\dagger$}  Department of Physics, The University of Michigan}
\centerline{\it Ann Arbor, MI 48109, USA}
\vspace{1cm}

\begin{abstract}

Writing the fully color dressed and graviton amplitudes,
respectively, as ${\bf A}=\langle C|A\rangle =\langle C|M|N\rangle$
and ${\bf A}_{gr}=\langle \tilde N|M|N\rangle$, where
$|A\rangle $ is a set of Kleiss-Kuijf
color ordered basis, $|N\rangle$, $|\tilde N\rangle $ and $|C\rangle$ are
the similarly ordered numerators and color coefficients,
we show that the propagator matrix $M$ has $(n-3)(n-3)!$ independent
eigenvectors $|\lambda ^0_j\rangle $ with zero eigenvalue, for $n$-particle processes.
The resulting equations $\langle \lambda ^0_j|A\rangle=0$ are
relations among the color ordered amplitudes.  The freedom
to shift $|N\rangle \to |N \rangle +\sum_j f_j|\lambda ^0_j\rangle$
and similarly for $|\tilde N\rangle $,
where $f_j$ are $(n-3)(n-3)!$ arbitrary functions, encodes
generalized gauge transformations.  They yield both BCJ amplitude
and KLT relations, when such freedom is accounted for.
Furthermore, $f_j$ can be promoted to the role of effective
Lagrangian vertices in the field operator space.

\end{abstract}

\end{titlepage}

%\end{center}

%\noindent

%\tableofcontents\eject

%%%%%%%%%%%%%%%%%%%%%%%%%%%%%%%%%%%%%%%%%%%%%%%%%%%%%%%%%%%%%%%%%%%%%%%%%%%%%%%%%%%%%%%%%%%%%%%%%%%%%%%%%%%%%%%%%%%%%%%%%%%%%%%%%%%%%%%%%%%%%%%%%%%%%%%%%%%%%
%%%%%%%%%%%%%%%%%%%%%%%%%%%%%%%%%%%%%%%%%%%%%%%%%%%%%%%%%%%%%%%%%%%%%%%%%%%%%%%%%%%%%%%%%%%%%%%%%%%%%%%%%%%%%%%%%%%%%%%%%%%%%%%%%%%%%%%%%%%%%%%%%%%%%%%%%%%%%
\section{Introduction and Summary}
%%%%%%%%%%%%%%%%%%%%%%%%%%%%%%%%%%%%%%%%%%%%%%%%%%%%%%%%%%%%%%%%%%%%%%%%%%%%%%%%%%%%%%%%%%%%%%%%%%%%%%%%%%%%%%%%%%%%%%%%%%%%%%%%%%%%%%%%%%%%%%%%%%%%%
$${}$$

In a series of papers, Bern and his collaborators \cite{Bern:2008qj}, \cite{Bern:2010ue}, \cite{Bern:2010yg}, and others
 using string theory insights \cite{BjerrumBohr:2009rd} or
through the use of the BCFW \cite{Britto:2005fq} recursion relations\footnote{The recursion relations found by Britto, Cachazo and Feng \cite{Britto:2004ap} among $n$-point color-ordered gauge theory tree-level amplitudes were proven in \cite{Britto:2005fq}, based on certain complex shifts of pairs of external gluon momenta, and the on the
behaviour of the tree level amplitude at large values of the complex shift parameter. In \cite{Vaman:2005dt}, the BCFW recursion relations were shown to originate in a set of identities, known as the largest time equation, which are obeyed in causal theories.
In \cite{Vaman:2008rr}, the BCFW shifts were extended to triple shifts of external gluon momenta to address recursion relations at one loop level. }  
\cite{Feng:2010my},
obtained some very interesting results regarding the number
of independent amplitudes ($(n-3)!$) for $n$ gluon scattering
and the relationship between gravitational and non-abelian
gauge amplitudes with the same external momenta.  The derivations in \cite{Bern:2008qj}
and arguments to arrive at these results rest heavily on duality
between intrinsic dynamics and color kinematics and on what is
termed generalized gauge transformations.  Thus, one
symbolically writes the color dressed $n$-gluon amplitude as \cite{DelDuca:1999rs}
\be
A^{(n)}=\sum _i {c_i n_i \over (\prod s_j)_i},       \label{acn}
\ee
where $c_i$ are color factors, $n_i$ numerators made of
kinematical invariants of the process, and $(\prod s_j)_i$
are appropriate products of inverse propagators.  The
conjecture, which has been subsequently proven,  is that
for channels with color factors satisfying
\be
c_i+c_j+c_k=0,                    
\ee
one can generate a set of numerators with only effective
triple vertices such that the corresponding ones also satisfy
\be
n_i+n_j+n_k=0.                    
\ee
Furthermore, Bern, Carrasco and Johansson (BCJ) \cite{Bern:2008qj}  
argue that there are $(n-3)(n-3)!$
degrees of arbitrariness in specifying these numerators,
which are made into generalized gauge transformations.
Basing on these, they showed that the number of independent
amplitudes for $n$ gluon scattering is $(n-3)!$ and that the
corresponding graviton scattering amplitude is  
\cite{Bern:2008qj},
\cite{Bern:2010ue},  
\cite{Bern:2010yg}
\be
A_{gr}^{(n)}=\sum _i {n_i \tilde n_i\over (\prod s_j)_i},    \label{grav_ampl}   
\ee
where $\tilde n_i$ are another copy of numerators due to
the same or a different gauge theory with
\be
\tilde c_i+\tilde c_j+\tilde c_k=0,
\ee
and
\be
\tilde n_i+\tilde n_j+\tilde n_k=0.
\ee
These relations between gravity amplitudes and gauge theory amplitudes, which are
inspired by the Kawai-Lewellen-Tye (KLT) relations \cite{klt} (for a more recent work see also \cite{Bern:1999bx}) and the BCJ relations have received a great deal of interest in the recent months:  \cite{Stieberger:2009hq}, \cite{Mafra:2009bz},
\cite{BjerrumBohr:2010zp},
\cite{BjerrumBohr:2010zs},
\cite{BjerrumBohr:2010ta},
\cite{Tye:2010dd},
\cite{Tye:2010kg},
and new identities (non-linear this time) were found among gauge theory amplitudes \cite{BjerrumBohr:2010zb}.

Since these results have profound consequences on how one
may perform gauge amplitude calculations and more importantly
how one may look at gravitational interaction, we would like
to investigate the origin of the gauge freedoms\footnote{In \cite{BjerrumBohr:2010zs}  the gauge freedom of Bern et al \cite{Bern:2008qj}, \cite{Bern:2010ue}  was interpreted as reparametrization invariance of the amplitude, such that it remains compatible with monodromy relations derived from string theory.}. Somewhat
surprisingly, we find that they come from the structure of the
$(n-2)!\times (n-2)! $ propagator matrix, when we work with the
Kleiss-Kuijf basis $A_i^{(n)}$ \cite{Kleiss:1988ne},
\be
M \sim ({1\over (\prod s_j)_i}),  
\ee
by which the color-ordered vector can be written as
\be
|A\rangle = M|N\rangle, \label{amn}
\ee
where $|N\rangle $ is a numerator vector with $n_i$ as its
entries
\be
|N\rangle=\begin{pmatrix}
n_1\\
n_2\\
\dots\\
n_{(n-2)!}
\end{pmatrix}.   
\ee
We shall find that the matrix $M$ has $(n-3)(n-3)!$ independent eigenvectors with zero eigenvalues
\be
M|\lambda ^0_j\rangle=0.  \label{null_vectors}
\ee
As a consequence, one can add
a linear combination of these null eigenvectors with the same
number of arbitrary functions $f_j$ as coefficients to the
numerator vector without changing the value of the color ordered scattering amplitude vector
\be
|A\rangle = M(|N\rangle +\sum_j f_j |\lambda ^0_j\rangle).
\ee
One can of course interpret this as effectively
making a transformation on the numerators
\be
|N \rangle \to |N'\rangle =|N\rangle +\sum_j f_j |\lambda ^0_j\rangle.
\ee
However, we should emphasize that we need not re-shuffle
entries in the original numerator vector $|N\rangle$ to accomplish
the change.  We just add a useful zero externally.

Another  immediate outcome is that from (\ref{amn}) and (\ref{null_vectors}),
one obtains $(n-3)(n-3)!$ independent relations among the color ordered
amplitudes
\be
\langle \lambda ^0_j|A\rangle = \langle \lambda ^0_j|M|N\rangle=0,
\ee
the precise form of which depends on our choice of
basis for $|A\rangle$ and $|\lambda ^0_j\rangle.$

To turn to gravity, let us be a bit more specific.  We
shall work in the Kleiss-Kuijf basis
\be
n_i=n(1,i_2, i_3,\dots n_{i_{n-1}},n).
\ee
for the $n$-particle color ordered amplitudes, in
which the labels denote the ordering of the
external legs.  There are obviously $(n-2)!$
entries for the numerator vector $|N\rangle_i=n_i$.  We shall label
the color factors in the same order and form
a vector
\be
\langle C|_i\equiv c_i=c(1,i_2, i_3,\dots n_{i_{n-1}},n),
\ee
in the same sequence as in $|N\rangle.$  It can be shown easily
that (\ref{acn}) for the color dressed amplitude becomes
\be
{\bf A}^{(n)}=<C|M|N>,  
\ee
and (\ref{grav_ampl}) for the $n$-graviton amplitude
\be
{\bf A}^{(n)}_{gr}=\langle \tilde N|M|N\rangle=\langle N|M|\tilde N\rangle,
\ee
as $M$ is symmetric.  Let us denote
\be
\sum_j f_j |\lambda ^0_j\rangle \equiv |\delta N\rangle.  
\ee
Clearly we have
\be
\delta{\bf  A}^{(n)}=\langle C|M |\delta N\rangle=0,\label{delta_A}
\ee
and
\be
\delta {\bf A}^{(n)}_{gr}=\langle \delta \tilde N|M|N\rangle
+\langle \tilde N|M|\delta N\rangle
+\langle \delta \tilde N|M|\delta N\rangle=0 \label{delta_A_grav}
\ee
term-wise because of (\ref{null_vectors}).  This is a statement of generalized
gauge invariance.  

In the recursive proof of squaring hypothesis of (\ref{grav_ampl}) for
any $n$, Bern et al used BCFW complexification to relate
the higher point numerator to the $\tilde n_i n_i$'s of lower
points.  They can differ by a set of gauge transformations
at every $z$-pole, which must manage to cancel to give the
whole amplitude  a gauge independent construct.    In
our formulation, the existence of null eigenvectors of $M$ depends
on overall energy momentum conservation and masslessness
of all external legs, which are respected by BCFW.  Thus  $M$
and $|\delta N \rangle$ can be extended to yield (\ref{delta_A}) and
(\ref{delta_A_grav}).
In fact, they are satisfied by the residues at every single $z$-pole
of $M$,
called channels, and moreover there is a unique $f_j$ at each
channel to effect the necessary  gauge transformation.   In short, we
find that our formulation is very natural for the study of the
issues involved.                      
                              
The plan of this article is as follows:  in the next two sections
(2 and 3) we shall specify the labeling $i$ of the numerators $n_i$
for four and five particle amplitudes in relation to the K-K basis.
A set of Jacobi identities for them will summarize the eventual
duality symmetry between $n_i$ and the associated color factors $c_i$.
The propagator matrix $M$ will be given, from which a set of
$(n-2)!$ null eigenvectors will be made explicit.  
The ranks of the null spaces for $n=4, 5$ will be found by
forming constraint matrices out of these null eigenvectors.  

In Section 4, we shall expose the source of generalized gauge
transformations, which is because of the existence of the null eigenvectors,
again explicitly for  $n=4, 5$.  We shall make use of specific
choices of the gauge functions $f_j$ and null eigenvectors $|\lambda ^0_j\rangle$
to show how relations between color ordered amplitudes and
graviton amplitudes are reached, and how in general the
structure of $f_j$ and $|\lambda ^0_j\rangle$ connects the roles
played by color Jacobi identities among $c_i$ for gauge amplitudes with those by $n_i$ for graviton amplitudes, to make gauge
invariance possible for every channel of the scattering amplitudes.

Some concluding remarks are made in Section 5, where we
also point out how natural it is to promote the gauge functions
$f_j$ into effective Lagrangian vertices in field space to make manifest
the semi-local nature of the necessary gauge transformations in
recursive constructions of $n_i$.

We compile an Appendix to discuss the six particle scattering
amplitudes.

%%%%%%%%%%%%%%%%%%%%%%%%%%%%%%%%%%%%%%%%%%%%%%%%%%%%%%%%%%%%%%%%%%%%%%%%%%%%%%%%%%%%%%%%%%%%%%%%%%%%%%%%%%%%%%%%%%%%%%%%%%%%%%%%%%%%%%%%%%%%%%%%%%%%%%%%%%%%%
%%%%%%%%%%%%%%%%%%%%%%%%%%%%%%%%%%%%%%%%%%%%%%%%%%%%%%%%%%%%%%%%%%%%%%%%%%%%%%%%%%%%%%%%%%%%%%%%%%%%%%%%%%%%%%%%%%%%%%%%%%%%%%%%%%%%%%%%%%%%%%%%%%%%%%%%%%%%%
\section{Constraints on Four Gluon  Amplitudes}
%%%%%%%%%%%%%%%%%%%%%%%%%%%%%%%%%%%%%%%%%%%%%%%%%%%%%%%%%%%%%%%%%%%%%%%%%%%%%%%%%%%%%%%%%%%%%%%%%%%%%%%%%%%%%%%%%%%%%%%%%%%%%%%%%%%%%%%%%%%%%%%%%%%%%%%%%%%%%%%%%%%%%%%%%%%%%%%%%%%%%%%%%%%%%%%%%%%%%%%%%%%%%%%%%%%%%%%%%%%%%%%%%%%%%%%%%%%%%%%%%%%%%%%%%%%%%%%%%%%%%%%%%%%%%%%%%%%%%%%%%%%%%%%%%%%%%%%%%%%%%%%%%%%%%%%%%%
$${}$$

In the Kleiss-Kuijf basis of A(1234) and A(1324), the amplitudes have simple poles in the $s_{12}, s_{14}$ and $s_{13}, s_{14}$ channels. Let us denote their corresponding numerators by $n_1=n(12;34), n_2=n(23;41)$ and $n_3=n(13;24), n_4=n(32;41)$, where the external gluons which share a vertex are denoted by pairs of indices not  separated by semi-columns. The cyclic order of the external gluons is indicated by the numerals 1 to 4 read clock-wise.
Due to Jacobi identities we have
\be
n(ij;kl)=-n(ji;kl),\qquad  n(ij;kl)+n(ki;jl)+n(jk;il)=0, \qquad n(ij;kl)=n(lk;ji)
\ee
which yield, in particular,
\be
n_4+n_3=n_1,\qquad n_2=-n_4.
\ee
This leads to the following relationship between the Kleiss-Kuijf amplitudes and the basis of independent numerators, $n_1, n_3$:
\be
\begin{pmatrix}
A(1234)\\
A(1324)
\end{pmatrix}=
\begin{pmatrix}
\frac{1}{s_{12}}+\frac{1}{s_{14}}&-\frac{1}{s_{14}}\\
-\frac{1}{s_{14}}&\frac{1}{s_{14}}+\frac{1}{s_{13}}
\end{pmatrix}
\begin{pmatrix}
n_1\\
n_3
\end{pmatrix},
\ee
or, in shorthand notation,
\be
|A^{(4)}\rangle =M^{(4)}|N^{(4)}\rangle,
\ee
where $|A\rangle$ is the Kleiss-Kuijf amplitudes column matrix, $M$ is the square matrix (and symmetric, in the chosen numerator basis) which encodes the pole structure of the amplitudes   and $|N\rangle $ is the numerator matrix.

The crucial observation made by Bern et al. in our language is that $M$ has rank $(n-3)!$ for a $(n-2)!$ set of independent $n$-point gluon amplitudes.
 A related observation (see \cite{BjerrumBohr:2009rd}, \cite{Feng:2010my})
 is that the Kleiss-Kuijf amplitudes matrix $A$ obeys a set of constraints which can be summarized by the following equation:
\bea
&&
\bigg(s_{i_2 i_3}+ s_{i_2 i_3}+\dots+ s_{i_2 i_{n-1}} +s_{i_2 n}\bigg)
A(1 i_2 i_3 \dots i_{n-1} n)\nn\\
&&+
\bigg(s_{i_2 i_3}+\dots +s_{i_2 i_{n-1}}+ s_{i_2 n}\bigg)
A(1 i_3 i_2 \dots i_{n-1} n)\nn\\
&& +\dots\nn\\
&&+\bigg( s_{i_2 i_{n-1}}+ s_{i_2 n}\bigg)
A(1 i_3 i_4\dots i_{n-2} i_{2} i_{n-1} n)\nn\\
&&+s_{i_2 n} A(1 i_3 i_4\dots i_{n-1} i_{2} n)  = 0\label{npoint_constraint}.
\eea
This means that there are eigenvectors $\langle \lambda ^0_j|$
of $M$ with zero eigenvalue
$$\langle \lambda ^0_j|M=0,$$
whose entries are $\bigg(s_{i_2 i_3} +s_{i_2 i_4}+\dots
+s_{i_2 i_{n-1}} +s_{i_2 n}\bigg)$ at  $(1 i_2 i_3 \dots i_{n-1} n)$,
.... , $s_{i_2 n}$ at  $(1 i_3 i_4\dots i_{n-1} i_{2} n)$.
There are $(n-2)!$ such vectors, but only $(n-3)(n-3)!$ of which
are linearly independent.  To see that we construct a constraint
matrix ${\cal C}$ by putting all these row null eigenvectors into a 
$(n-2)! \times (n-2)!$ matrix with the property
\be
{\cal C}|A\rangle={\cal C}M|N\rangle=0,
\ee
which will be shown to have rank $(n-3)!$.

In the case of the four-point amplitude, it is easy to show that
$M^{(4)}$ has rank 1.  Equivalently, we have
\be
\langle \lambda^0_1|=\langle s_{23}+s_{24} \; \;s_{24}|,
\ee
\be
\langle \lambda^0_2|=\langle s_{34}  \;\; s_{23} +s_{34}|,
\ee
and the constraint matrix is
\be
{\cal C}^{(4)}=\begin{pmatrix} s_{23}+s_{24} & s_{24}\\s_{34}&s_{23}+s_{34}\end{pmatrix}.
\ee
The two constraints obeyed by the four-point Kleiss-Kuijf gluon amplitudes are not actually independent of each other. The rank of the constraint matrix 
${\cal C}^{(4)}$ is 1, which can be seen by noticing that, due to momentum conservation and the on-shellness of the external gluon momenta, the sum of the rows of the matrix ${\cal C}^{(4)}$ is 0.
This feature of the constraint matrix continues to hold for all $n$-point amplitudes.
 
%%%%%%%%%%%%%%%%%%%%%%%%%%%%%%%%%%%%%%%%%%%%%%%%%%%%%%%%%%%%%%%%%%%%%%%%%%%%%%%%%%%%%%%%%%%%%%%%%%%%%%%%%%%%%%%%%%%%%%%%%%%%%%%%%%%%%%%%%%%%%%%%%%%%%%%%%%%%%
\section{Constraints on Five Gluon Amplitudes}
%%%%%%%%%%%%%%%%%%%%%%%%%%%%%%%%%%%%%%%%%%%%%%%%%%%%%%%%%%%%%%%%%%%%%%%%%%%%%%%%%%%%%%%%%%%%%%%%%%%%%%%%%%%%%%%%%%%%%%%%%%%%%%%%%%%%%%%%%%%%%%%%%%%%%%%%%%%%%
$${}$$

The Kleiss-Kuijf basis of gluon amplitudes can be chosen to be: $A(12345),$ $A(14325),$ $A(13425),$ $A(12435),$ $A(14235),$ $A(13245)$\footnote{For an early explicit calcualtion of five and six-gluon tree-level amplitudes see \cite{Berends:1987cv}.}. Each of these amplitudes has simple poles in the various kinematic invariants. We will denote the numerators associated with these poles as follows:
\bea
&&n_1=n(12;3;45),\qquad n_2=n(23;4;51),\qquad n_3=-n(12;5;34),\qquad
n_4=n(23;1;45),\nn\\&& n_5=n(51;2;34),\qquad
n_6=n(14;3;25),\qquad n_7=n(32;5;14), \qquad n_8=n(25;1;43),\nn\\
&&
n_9=n(13;4;25),\qquad n_{10}=-n(13;5;42),\qquad
n_{11}=n(51;3;42),\qquad n_{12}=n(12;4;35),\nn\\
&& n_{13}=n(35;1;24),\qquad
n_{14}=n(14;2;35),\qquad n_{15}=n(13;2;45).
\eea
Our notation for the numerators associated with a certain tree level Feynman diagram is as follows: we specify the external gluons (with certain momenta and helicities) by numerals; those  external gluons which are indicated by numerals not separated by  semi-columns share a common vertex; the remaining external gluon which joins some internal lines in a vertex is indicated by a separate numeral; all external gluon lines are written in a clock-wise order. Of course, due to Jacobi identities at the vertex on the internal lines, the numerators  obey straightforward identities associated with flipping the orientation of the external line above or below the internal lines. For example:
\be
n(13;5;42)=-n(13;42;5).
\ee
Similarly, flipping the order of the external gluons with a common vertex brings a minus sign, e.g.:
\be
n(12;3;45)=-n(21;3;45)=-n(12;3;54).
\ee
In addition we have relations between the fifteen numerators which also follow from the use of Jacobi identities:
\bea
&&n_1+n_4=n_{15},\qquad n_1-n_3=n_{12},\qquad n_6-n_7=n_{14},\qquad
n_2+n_4=n_7,\nn\\
&&n_9-n_{15}=n_{10},\qquad n_6=n_8=n_9,\qquad -n_3+n_5=n_8,\qquad
n_2-n_5=n_{11},\nn\\
&&n_{11}-n_{10}=n_{13}.
\eea
In more generality, the numerators obey Jacobi-type identities
\bea
&&n(ij;k;lm)=-n(ji;k;lm), \nn\\
&&n(ij;k;lm)+n(ki;j;lm)+n(jk;i;lm)=0,
\eea
in addition to the time-reverse identity
\be
\qquad n(ml;k;ji)=-n(ij;k;lm).
\ee
Eliminating the other numerators in favor of $n_1, n_6, n_9, n_{12}, n_{14}, n_{15}$, we can express the Kleiss-Kuijf amplitudes in terms of the numerators
\be
\begin{pmatrix}
A(12345)\\
A(14325)\\
A(13425)\\
A(12435)\\
A(14235)\\
A(13245)%\rangle
\end{pmatrix}
=M^{(5)}
\begin{pmatrix}
n_1\\
n_6\\
n_9\\
n_{12}\\
n_{14}\\
n_{15}%\rangle,
\end{pmatrix}
\ee
where $M^{(5)}$ is given by
\be
\!\!\!\!\!\!\!\!\!\!\begin{pmatrix}
\frac{1}{s_{12}s_{45}}+\frac{1}{s_{15}s_{34}}&\frac{1}{s_{15}s_{34}}+\frac{1}{s_{23}s_{15}}&-\frac{1}{s_{15}s_{34}}&-\frac{1}{s_{15}s_{34}}-\frac1{s_{12}s_{34}}&
-\frac{1}{s_{23}s_{15}}&-\frac{1}{s_{23}s_{45}}-\frac{1}{s_{23}s_{15}}\\
+\frac{1}{s_{23}s_{15}}+\frac{1}{s_{12}s_{34}}& & & & &\\
+\frac{1}{s_{23}s_{45}}& & & & & \\
\\
\frac{1}{s_{15}s_{34}} +\frac{1}{s_{15}s_{23}}&\frac{1}{s_{14}s_{25}}
+\frac{1}{s_{14}s_{23}} &-\frac{1}{s_{15}s_{34}} -\frac{1}{s_{34}s_{25}}&
-\frac{1}{s_{15}s_{34}} & -\frac{1}{s_{14}s_{23}}-\frac{1}{s_{15}s_{23}}&
-\frac{1}{s_{15}s_{23}}\\
&+\frac{1}{s_{15}s_{23}}+\frac{1}{s_{15}s_{34}}& & & &\\
&+\frac{1}{s_{34}s_{25}}& & & &\\
\\
-\frac{1}{s_{15}s_{34}}& -\frac{1}{s_{15}s_{34}}-\frac{1}{s_{34}s_{25}}&
\frac{1}{s_{13}s_{25}}+\frac{1}{s_{13}s_{24}} & \frac{1}{s_{15}s_{24}}+
\frac{1}{s_{15}s_{34}}&-\frac{1}{s_{15}s_{24}} &-\frac{1}{s_{13}s_{24}}
-\frac{1}{s_{15}s_{24}}\\
&&+\frac{1}{s_{15}s_{24}}+\frac{1}{s_{15}s_{34}}& & &\\
&&+\frac{1}{s_{34}s_{25}}\\
\\
-\frac{1}{s_{12}s_{34}}-\frac{1}{s_{15}s_{34}}&-\frac{1}{s_{15}s_{34}} &
\frac{1}{s_{15}s_{34}}+\frac{1}{s_{15}s_{24}} & \frac{1}{s_{12}s_{35}}+
\frac{1}{s_{12}s_{34}}&-\frac{1}{s_{15}s_{24}}-\frac{1}{s_{24}s_{35}} &
-\frac{1}{s_{15}s_{24}}\\
&&&+\frac{1}{s_{15}s_{34}}+\frac{1}{s_{15}s_{24}}&&\\
&&&+\frac{1}{s_{24}s_{35}}&&\\
\\
-\frac{1}{s_{15}s_{23}} & -\frac{1}{s_{14}s_{23}}-\frac{1}{s_{15}s_{23}}&
-\frac{1}{s_{15}s_{24}} &-\frac{1}{s_{15}s_{24}}-\frac{1}{s_{24}s_{35}} &
\frac{1}{s_{14}s_{35}}+\frac{1}{s_{14}s_{23}} &\frac{1}{s_{15}s_{23}}+\frac{1}{s_{15}s_{24}}\\
&&&&+\frac{1}{s_{15}s_{23}}+\frac{1}{s_{15}s_{24}}&\\
&&&&+\frac{1}{s_{24}s_{35}}&\\
\\
-\frac{1}{s_{23}s_{45}}-\frac{1}{s_{23}s_{15}}& -\frac{1}{s_{15}s_{23}}&
-\frac{1}{s_{13}s_{24}}-\frac{1}{s_{15}s_{24}}&-\frac{1}{s_{15}s_{24}} &
\frac{1}{s_{15}s_{23}}+\frac{1}{s_{15}s_{24}} &\frac{1}{s_{13}s_{45}}
+\frac{1}{s_{13}s_{24}}\\
&&&&&+\frac{1}{s_{15}s_{24}}+\frac{1}{s_{15}s_{23}}\\
&&&&&+\frac{1}{s_{23}s_{45}}
\end{pmatrix}.
\ee
The symmetric matrix $M^{(5)}$ used to express the Kleiss-Kuijf five-point amplitudes
in the basis of the independent numerators has rank 2 (and thus equal to (n-3)! for $n$-point amplitudes).

This is also reflected by the existence of a constraint matrix such that
${\cal C}^{(5)}|A^{(5)}\rangle=0$ and ${\cal C}^{(5)}M^{(5)}=0$, where ${\cal C}^{(5)}$ is obtained from specializing the constraint equations (\ref{npoint_constraint}) to the case of the chosen five-point Kleiss-Kuijf amplitude basis:
\be
\!\!\!\!\!\!\!\!\!%C=
\begin{pmatrix}
s_{23}+s_{24}+s_{25}&0&s_{25}&0&0&s_{24}+s_{25}\\
0&s_{24}+s_{34}+s_{45}&s_{24}+s_{45}&0&0&s_{45}\\
0&s_{23}+s_{35}&s_{34}+s_{23}+s_{35}&0&s_{35}&0\\
0&s_{25}&0&s_{24}+s_{23}+s_{25}&s_{23}+s_{25}&0\\
s_{45}&0&0&s_{34}+s_{45}&s_{24}+s_{34}+s_{45}&0\\
s_{34}+s_{35}&0&0&s_{35}&0&s_{23}+s_{34}+s_{35}
\end{pmatrix}.
\ee
The constraint matrix ${\cal C}^{(5)}$ has rank 4. In other words, not all constraint equations are independent, and so only four linear constraints can be used to express the six Kleiss-Kuijf five-point amplitudes in terms of each other, leaving us with two independent Kleiss-Kuijf amplitudes (that is, the number of independent color-ordered amplitudes is $(n-3)!$). The null eigenvectors of matrix ${\cal C}^{(5)}$ (the elements of the kernel of ${\cal C}^{(5)}$) can be constructed by hand:
\bea
%\begin{pmatrix}
\bigg|-\frac{s_{13}(s_{34}+s_{45})}{s_{34}s_{12}}\ \
-\frac{s_{45}s_{13}}{s_{34}s_{25}}\ \
-\frac{s_{45}(s_{23}+s_{35})}{s_{34}s_{25}}\ \
-\frac{s_{45}s_{13}}{s_{34}s_{12}}\ \ 0\ \ 1\big\rangle\nn\\
%\end{pmatrix},\qquad
%\begin{pmatrix}
\bigg|-\frac{s_{35}s_{14}}{s_{34}s_{12}}\ \
-\frac{s_{35}(s_{24}+s_{45})}{s_{34}s_{25}}\ \
-\frac{s_{35}s_{14}}{s_{34}s_{25}}\ \
-\frac{s_{14}(s_{34}+s_{35})}{s_{34}s_{12}}\ \
1\ \ 0\big\rangle.
%\end{pmatrix}
\eea
 Of course, it is not a coincidence that the dimensionality of the null space of the matrix ${\cal C}^{(5)}$ and the rank of $M^{(5)}$ are the same.
The null eigenvectors of $M^{(5)}$ can be found by solving $M^{(5)}|\lambda^0\rangle=0$, which is identical with
\be
\langle \lambda^0|M^{(5)}=0,
\ee
since $M^{(5)}$ is symmetric. The existence of $|\lambda^0\rangle$ implies that the amplitudes must obey a set of linear relations which follow from
\be
\langle \lambda^0|A\rangle=0.
\ee
Fortunately, once one of the null eigenvectors is found, the others are obtained by permutations of indices. Let us assume that there is such a null vector of the form $|\lambda^0_1\rangle = |a_1, a_2, a_3, 0,0,0\rangle$. Solving $M^{(5)}|\lambda^0_1\rangle=0$ yields
\be
|\lambda^0_1\rangle= |s_{12}s_{45},-s_{14}(s_{24}+s_{25}),s_{13}s_{24},0,0,0\rangle.
\ee
The next null vector is of the form $|\lambda^0_2\rangle=|b_1 ,b_2,0,b_4,0,0\rangle$. We notice that $a_3=s_{13}s_{24}$ is the entry with indices 13425 in the Kleiss-Kuijf basis, and that $b_4$ corresponds to the entry 12435=53421. Therefore this null vector can be obtained from $|\lambda^0_1\rangle$ by interchanging the indices $1$ and $5$. This gives
\bea
&&a_1=s_{12}s_{45} , \qquad 12345\longleftrightarrow 52341=14325 \qquad b_2=s_{25}s_{14}\nn\\
&&a_2= -s_{14}(s_{24}+s_{25}),\qquad 14325\longleftrightarrow 54321=12345 \qquad b_1=-s_{45}(s_{24}+s_{12})\nn\\
&&a_3=s_{13}s_{24},\qquad 13425\longleftrightarrow 12435\qquad b_4=s_{35}s_{24},
\eea
and so
\be
|\lambda^0_2\rangle = |-s_{45}(s_{12}+s_{24}),s_{14}s_{25},0,s_{35}s_{24},0,0\rangle.
\ee
Similar arguments lead to the identification of the other two null eigenvectors:
\bea
&&|\lambda^0_3\rangle=|s_{12}s_{45},-s_{25}(s_{14}+s_{24}),0,0,s_{35}s_{24},0\rangle\\
&&|\lambda^0_4\rangle= | -s_{12}(s_{24}+s_{45}),s_{14}s_{25},0,0,0,s_{13}s_{24}\rangle.
\eea
This set of null eigenvectors will be used in the next section.

%%%%%%%%%%%%%%%%%%%%%%%%%%%%%%%%%%%%%%%%%%%%%%%%%%%%%%%%%%%%%%%%%%
%%%%%%%%%%%%%%%%%%%%%%%%%%%%%%%%%%%%%%%%%%%%%%%%%%%%%%%%%%%%%%%%%%
\section {Gauge Freedom}
%%%%%%%%%%%%%%%%%%%%%%%%%%%%%%%%%%%%%%%%%%%%%%%%%%%%%%%%%%%%%%%%%%%%%%%%%%%%%%%%%%%%%%%%%%%%%%%%%%%%%%%%%%%%%%%%%%%%%%%%%%%%%%%%%%%%%%%%%%%%%%%%%%%%%%%%%%%%%%%%%%%%%%%%%%%%%%%%%%%%%%%%%%%%%%%%%%%%%%%%%%%%%%%%%%%%%%%%%%%%%%%%%%%%%%%%%%%%%%%%%%%%%%%%%%%%%%%%
$${}$$

We have discussed the structure of the null eigenvectors of $M$
in the last section.  A natural question may be what about the
eigenvectors with non-zero eigenvalues.  As one has
\be
M=\sum_i \lambda_i |\lambda_i\rangle \langle\lambda_i|,%(4.1)
\ee
one may suspect that these eigenvalues and vectors with
$\lambda_i\neq 0$ may be essential in understanding  the
structure of the space of tree level amplitudes.  For example,
what linear combinations of the numerators should one be
interested in to yield the amplitudes.  However, this intuitive
reasoning is not particularly rewarding, because both
$\lambda _i$ and $|\lambda _i\rangle$ can have non-analytic
dependence on kinematical invariants $s_i$.  They must
of course cancel out in any expression of physical significance, but
their presence strongly indicates that we should seek a
different path.  

It turns out that the null eigenvectors are there to
yield most of the information of this kind that we want to
obtain.  As we pointed out, we are free to add $\sum f_i
|\lambda ^0_i\rangle $ to $|N\rangle $.  By adjusting these gauge functions
$f_i$, we are able to isolate the combinations of $n_i$ which
are relevant to the amplitudes and obtain relations amongst
them.  Furthermore, we can construct higher point $n_i$
recursively via BCFW continuation with lower point ones.
The difference as it turns out is a gauge transformation.
This opens up a new way to construct recursively an
effective Lagrangian to study loops.  What has to be
done is of course to extend the on-shell gauge freedom
to off-shell, which will be briefly touched on in the Concluding Remarks section.

\subsection{ Four Particle Amplitudes}

To discuss gauge freedom, we begin by rewriting the
color-dressed amplitude.  We shall drop a common
factor proportional to powers of the coupling constant.
Thus,
\be
{\bf A}^{(4)}={n(12;34)c(1234)\over s_{12}}
                +{n(13;42)c(1342)\over s_{13}}
                +{n(14;23)c(1423)\over s_{14}},  % (4.2)
\ee
where the color factors are
\be
c(1234)=\sum_g f_{a_1a_2g}f_{a_3a_4g}, \ \  etc.,  %(4.3)
\ee
$f_{a_ia_ja_k}$ being the totally antisymmetric structure constants of the symmetry group.
The antisymmetry and the Jacobi identity of these constants
\be
\sum_g \big( f_{a_1a_2g}f_{a_3a_4g}+f_{a_2a_3g}f_{a_1a_4g}
                      +f_{a_3a_1g}f_{a_2a_4g}\big)=0,      %(4.4)
\ee
are written into
\be
c(jikl)=-c(ijkl), \qquad c(lkji)=c(ijkl), \qquad
c(ijkl)+c(jkil)+c(kijl)=0,% (4.5)
\ee
just as the $n(ij;kl)$'s.  Simple algebra then
gives
\be
{\bf A}^{(4)}=\langle C^{(4)}|M^{(4)}|N^{(4)}\rangle,  %(4.6)
\ee
in which
\be
\langle C^{(4)}|=\langle c(1234) \;\; c(1324) |, \qquad  
\langle N^{(4)}|=\langle n(12;34) \;\; n(13;24) |,
%(4.7)
\ee
and $M^{(4)}$ is given in eq. (2.3).

The four graviton amplitude, according to the the squaring
hypothesis, can be rearranged into
\bea
 {\bf A}_{gr}^{(4)}&=&{\tilde n(12;34)n(12;34)\over s_{12}}
                +{\tilde n(13;42)n(13;42)\over s_{13}}
                +{\tilde n(14;23)n(14;23)\over s_{14}}
                 \nn\\
& =&\langle \tilde N^{(4)}|M^{(4)}|N^{(4)}\rangle,
                  %(4.8)
\eea
with                   
\be
\langle \tilde N^{(4)}|=\langle
\tilde n(12;34) \ \ \tilde n(13;24) |.   %(4.9)
\ee
This rearrangement is again made possible by using the Jacobi
identities of the $n$'s and $\tilde n$'s.

As already pointed out, $M^{(4)}$ has one null eigenvector.  
We may choose, for example
\be
\langle \lambda ^0|=\langle -s_{12} \ \ s_{13}|.   %(4.10)
\ee
Then, one can add
\be
|\delta N^{(4)}\rangle =f|\lambda ^0\rangle
% (4.11)
\ee
to the color-ordered amplitude equation
\be
|A^{(4)}\rangle =M^{(4)}|N^{(4)}\rangle    \qquad \to \qquad  
|A^{(4)}\rangle =M^{(4)}(|N^{(4)}\rangle+|\delta N^{(4)}\rangle),   
%(4.12)
\ee
where $f$ is an arbitrary function in general.  However,
depending on what issues we are interested in, we
should choose $|\lambda ^0\rangle$ and $f$ accordingly.
For example, if we choose
\be
f=-{n(13;24)\over s_{13}},   %(4.13)
\ee
then
\be
\langle N^{(4)}|+\langle \delta N^{(4)}|=\langle n_1'=n(12;34)+{s_{12}\over
s_{13}}n(13;24)  \;\; 0  |,    %(4.14)   
\ee
which gives
\be
A(1234)=-{s_{13}\over s_{12}s_{14}}n_1', \qquad
A(1324)=-{1\over s_{14}}n_1', %(4.15)
\ee
or
\be
A(1234)={s_{13}\over s_{12}}A(1324),  %(4.16)
\ee
which implies that there is only one independent color-ordered amplitude.

Note that once we have fixed a particular set of numerators
as basis, in our case $n(12;34)$ and $n(13;24)$, all others must be
expressed in terms of them in making gauge transformations.
Therefore, since
\be
n(13;42)=-n(13;24), \qquad n(14;23)=n(13;24)-n(12;34), %(4.17)
\ee
we infer that the shifts of eq. (4.11)
\be
\delta n(12;34)=-s_{12}f, \qquad \delta n(13;24)=s_{13}f,  %(4.18)
\ee
should give also
\be
\delta n(13;42)=-s_{13}f, \qquad
\delta n(14;23)=-s_{14}f.  %(4.19)
\ee

It is illuminating to check gauge invariance of the amplitudes by
substituting eqs. (4.18-4.19) directly into the variation of eq. (4.2)
\be
\delta {\bf A}^{(4)}= -f(c(1234)+c(1342)+c(1423))=0,  %(4.20)
\ee
due to Jacobi identity of the last three indices.  Similarly,
\bea
\delta {\bf A}_{gr}^{(4)}&=& -f(\tilde n(12;34)+\tilde n(13;42)+\tilde n(14;23))
                                    \nn\\ & -&\tilde f(n(12;34)+n(13;42)+n(14;23))
                                     \nn\\ &+&f\tilde f (s_{12}+s_{13}+s_{14}),% (4.21)
\eea
upon making a similar gauge transformation on $\tilde n$'s.  Each
term in eq. (4.21) vanishes individually, because of the Jacobi identities of $\tilde n$'s, $n$'s and momentum conservation, respectively.  
They are made possible, because the kinematical invariants
$s_{12}, s_{13}$, and $s_{14}$ so aptly generated by the
null vector and then multiplied
to $f$ in eqs. (4.18-4.19) cancel out the propagators in
eqs. (4.2) and (4.8).  This is a general
feature for higher point amplitudes also.

%%%%%%%%%%%%%%%%%%%%%%%%%%%%%%%%%%%%%%%%%%%%%%%%%%%%%%%%%%%%%%%%%%%%%%%%%%
\subsection{ Five Particle Amplitudes}
%%%%%%%%%%%%%%%%%%%%%%%%%%%%%%%%%%%%%%%%%%%%%%%%%%%%%%%%%%%%%%%%%%%%%%%%%%

The color-dressed five gluon amplitude is
\bea
{\bf A}^{(5)}&=& {c(12345)n(12;3;45)\over s_{12}s_{45}}+
{c(23451)n(23;4;51)\over s_{23}s_{51}}+
{c(34512)n(34;5;12)\over s_{34}s_{12}}\nn\\ &+&
{c(45123)n(45;1;23)\over s_{45}s_{23}}+
{c(51234)n(51;2;34)\over s_{51}s_{34}}+                                     
{c(14325)n(14;3;25)\over s_{14}s_{25}}\nn\\ & +&
{c(32514)n(32;5;14)\over s_{32}s_{14}}+
{c(25143)n(25;1;43)\over s_{25}s_{43}}+        
{c(13425)n(13;4;25)\over s_{13}s_{25}}\nn\\ & +    &                        
{c(42513)n(42;5;13)\over s_{42}s_{13}}+                              
{c(51342)n(51;3;42)\over s_{51}s_{42}}+
{c(12435)n(12;4;35)\over s_{12}s_{35}}\nn\\ & +&
{c(35124)n(35;1;24)\over s_{35}s_{24}}+
{c(14235)n(14;2;35)\over s_{14}s_{35}}+
{c(13245)n(13;2;45)\over s_{13}s_{45}},%(4.22)
\eea
where the color factors are
\be
c(ijklm)=\sum _{b \ c}f_{a_i  a_j b}f_{ba_kc}f_{ca_la_m}, %(4.23)
\ee
which satisfy the same identities as $n(ij;k;lm)$
\bea
&&c(jiklm)=-c(ijklm), \qquad c(mlkji)=-c(ijklm), \nn\\
&&c(ijklm)+c(jkilm)+c(kijlm)=0.
 % (4.24)
\eea
Using these, we can relate the color-ordered amplitudes of
eq.(3.6) $|A^{(5)}\rangle=M^{(5)}|N^{(5)}\rangle$ to color-dressed
\be
{\bf A}^{(5)}=\langle C^{(5)}|A^{(5)}\rangle =\langle C^{(5)}|M^{(5)}|N^{(5)}\rangle,  %(4.25)
\ee
with
\be
\langle C^{(5)}|=\langle c(12345) \;\; c(14325) \;\; c(13425) \;\;
                  c(12435) \;\; c(14235) \;\; c(13245)|. %(4.26)
\ee
Similarly, the five graviton amplitude is
\be
{\bf A}^{(5)}_{gr}=\langle \tilde N^{(5)}|A^5\rangle =\langle \tilde N^{(5)}|M^{(5)}|N^{(5)}\rangle ,%(4.27)
\ee
with the same ordering of indices for $\tilde N^{(5)}$ as in $N^{(5)}$,
namely            
\be      
\langle \tilde N^{(5)}|=\langle \tilde n(12;3;45) \;\; \tilde n(14;3;25) \;\;
\tilde n(13;4;25) \;\; \tilde n(12;4;35) \;\; \tilde n(14;2;35) \;\; \tilde n(13;2;45)|.
%(4.28)
 \ee                 
 In a recursive proof of the squaring hypothesis to go from
 $n=4$ to $n-5$ graviton amplitudes, Bern et al explicitly obtained
 a gauge arbitrariness of the numerators.  We would like to
 show that the form of the gauge transformations which
 warrants gauge invariance {\it at each channel}
 is predicated by the null eigenvectors of
 $M^{(5)}$.  For this purpose, it is convenient to take the four
 independent null eigenvectors as in eqs. (3.13, 3.15,
 3.16 and 3.17).
% \bea
% && \langle \lambda _1^0|= \langle s_{12}s_{45} \ -s_{14}(s_{24}+s_{25})
% \ s_{13}s_{24} \ \ 0 \ \ 0 \ \ 0)|,\nn\\ &&
% \langle \lambda _2^0|= \langle -s_{45}(s_{12}+s_{24}) \ s_{14}s_{25} \  \ 0 \ \
% s_{24}s_{35} \ \ 0 \ \ 0 )|,\nn\\ &&
%\langle \lambda _3^0|= \langle s_{12}s_{45} \ -s_{25}(s_{14}+s_{24}) \ \ 0 \ \ 0
% \ s_{24}s_{35} \ \ 0 )|,\nn\\ &&
%  \langle \lambda _4^0|= \langle -s_{12}(s_{24}+s_{45}) \ s_{14}s_{25} \ \  0 \ \
 % 0 \ \ 0 \ \ s_{13}s_{24}  )|.%}(4.28)
%\eea
 
 The induced shifts on the numerators corresponding to
  \be
|\delta N^5\rangle=\sum _{i=1}^4 f_i |\lambda _i^0\rangle %(4.29)
 \ee
 are
\bea
&& \delta n(12;3;45)=s_{12}s_{45} (f_1-f_2+f_3-f_4)
                                                   -s_{24}(s_{45}f_2+s_{12}f_4), \nn\\ &&
\delta n(14;3;25)=-s_{14}s_{25} (f_1-f_2+f_3-f_4)
                                                   -s_{24}(s_{14}f_1+s_{25}f_3), \nn\\ &&
\delta n(13;4;25)=s_{13}s_{24}f_1, \nn\\ &&                 
\delta n(12;4;35)=s_{24}s_{35}f_2,  \nn\\ &&               
\delta n(14;2;35)=s_{24}s_{35}f_3.  \nn\\ &&                  
\delta n(13;2;45)=s_{13}s_{24}f_4.%(4.30)                 
    \eea              
Let us consider BCFW complexification of the amplitudes so that
we can relate five-point graviton amplitude to three and four-point amplitudes where squaring hypothesis is known to be true.  We shift
the momenta $p_1$ and $p_5$ according to
\be
|\hat 1]=|1]-z|5],     \ \ \ |\hat 5\rangle  =|5\rangle +z|1\rangle ,
%(4.31)
\ee
which will give rise to poles in the complex z-plane due
to the vanishing of $\hat s_{12}, \ \hat s_{13}$, $\hat s_{14},$
$ \hat s_{25}, \ \hat s_{35},$ and $\hat s_{45}.$  We call these channels
and examine the changes due to $\delta n$ on the
color-dressed gluon amplitude and the graviton amplitude at
each channel.  For $\hat s_{12}=0$, eq.(4.22) informs us that
the change is
\bea
\delta {\bf A}^{(5)} &=&{1\over s_{12}}\bigg[ {c(12345)
\delta \hat n(12;3;45)\over \hat s_{45}}
    + {c(34512)\delta \hat n(34;5;12)\over s_{34}}               
    +{c(12435)\delta \hat n(12;4;35)\over \hat s_{35}} \bigg]
 \nn\\ &&           
    =  {1\over s_{12}}\bigg[ {c(12345)\delta \hat n(12;3;45)
   \over \hat s_{45}}+ {c(12534)\delta \hat n(12;5;34)\over s_{34}}                   +{c(12453)\delta \hat n(12;4;53)\over \hat s_{35}} \bigg].\nn\\
%(4.32)  
\eea        
According to eq.(4.30), we should have
\bea
&& \delta \hat n(12;3;45)=-s_{24}\hat s_{45}\hat f_2,\nn\\&&
\delta \hat n(12;4;53)=-\delta \hat n(12;4;35)=-s_{24}\hat s_{35}\hat f_2,
\nn\\  &&
\delta \hat n(12;5;34)=-\delta \hat n(12;3;45)-\delta \hat n(12;4;53)\nn\\ &&                  
                  \ \ \ \ \ \ \ \ \ \ \ \ \ \ \ \ =s_{24}(\hat s_{45}+\hat s_{35})\hat f_2=-s_{24}s_{34}\hat f_2.%(4.33)
\eea
Putting these into eq.(4.32), we have
\bea
\delta {\bf A}^{(5)} =-{s_{24} \over s_{12}}
      \big[  c(12345)+   c(12534)+   c(12453)\big]\hat f_2 =0 %(4.34)
\eea
in view of a color Jacobi identity.  

As for the five graviton amplitude, we replace the $c(ijklm)$
in eq.(4.22) with $\tilde n(ij;k;lm)$, whose variation is:
 \bea
\delta {\bf A}_{gr}^{(5)} &=&         
    {1\over s_{12}}\bigg[ {\tilde n(12;3;45)\delta \hat n(12;3;45)
     \over \hat s_{45}}
    + {\tilde n(12;5;34)\delta \hat n(12;5;34)\over s_{34}}               
    +{\tilde n(12;4;53)\delta \hat n(12;4;53)\over \hat s_{35}}  \nn\\ &&        
    + {\delta \hat {\tilde n}(12;34;5) \hat n(12;3;45)\over \hat s_{45}}
    + {\delta \hat {\tilde n}(12;5;34) \hat n(12;5;34)\over s_{34}}               
    + {\delta \hat {\tilde n}(12;4;53) \hat n(12;4;53)\over \hat s_{35}} \nn\\ &&
  + {\delta \hat {\tilde n}(12;3;45) \delta \hat n(12;3;45)\over \hat s_{45}}
    + {\delta \hat {\tilde n}(12;5;34) \delta \hat n(12;5;34)\over s_{34}}               
    +{\delta \hat {\tilde n}(12;4;53) \delta \hat n(12;4;53)\over \hat s_       
       {35}}\bigg] \nn\\ &&
    ={1\over s_{12}}\big[-s_{24}\big(\hat {\tilde n}(12;3;45)
                                              + \hat {\tilde n}(12;5;34)
                                              +\hat {\tilde n}(12;4;53)\big)\hat f_2 \nn\\ &&
                                    \ \ \ \ -s_{24}\big(\hat n(12;3;45)+ \hat n(12;5;34)
                                              +\hat n(12;4;53)\big) \hat {\tilde f}_2 \nn\\ &&
                                    \ \ \ \ +s_{24}^2( s_{34}+\hat s_{35}+\hat s_{45})
                                    \hat f_2\hat {\tilde f}_2\big],
%(4.35)
\eea
 whereby each term vanishes on account of Jacobi identities and
 momentum conservation, just as in the four particle case.
 
 Let us briefly run through the other channels. For $\hat s_{13}=0$,
 the relevant changes in numerators are
 \bea
 &&\delta \hat n(13;4;25)=\hat s_{13} s_{24}\hat f_1=0, \nn\\ &&
 \delta  \hat n(13;2;45)=\hat s_{13}s_{24}\hat f_4=0,\nn\\ &&
 \delta \hat n(42;5;13) =- \delta  \hat n(13;2;45)+\delta \hat n(13;4;25)
 =0,%(4.36)
\eea
 for $\hat f_1$ and $\hat f_4$ which are chosen not to blow up
 at $\hat s_{13}=0.$  In other words, a gauge transformation is
not needed.

For $\hat s_{14}=0$, eqs.(4.2), (4.9) and (4.30) instruct that we
need the gauge shifts
\bea
&&\delta \hat n(14;3;25)=-s_{24}\hat s_{25}\hat f_3, \nn\\ &&
 \delta \hat n(14;2;35)=s_{24}\hat s_{35}\hat f_3.\nn\\ &&
 \delta \hat n(32;5;14)=-\delta \hat n(14;2;35)+\delta \hat n(14;3;25)\nn\\ &&
 \ \ \ \ \ \ \ \ \ \ \ \ \ \ \ \ =-s_{24}(\hat s_{35}+\hat s_{25})\hat f_3
 =s_{24}s_{23}\hat f_3.%(4.37)
\eea
 The  factors $\hat s_{25}, \hat s_{35}$ and $s_{23}$
 are there to cancel out the matching propagators in
 $\delta  {\bf A}^{(5)}$ and $\delta  {\bf A}^{(5)}_{gr}$ so that we
 can use Jacobi identities and momentum conservation
 to obtain gauge invariance.  
 
 Along the same vein, for $\hat s_{25}
 =0$, the gauge shifts are proportional to $\hat f_1$ with the
 matching factors $\hat s_{14}, \hat s_{13}$ and $s_{34}$
 to facilitate gauge invariance in that channel.  As for
 $\hat s_{35}=0$, all the necessary shifts in $n_i$ are
 proportional to $\hat s_{35}$ and therefore we don't
 have a gauge shift.
 
 The last channel $\hat s_{45}=0$ deserves a bit more
 exposition, because it was explicitly worked out by
 Bern et al when they  constructed $n_i$ for five particles
 from those for three and four.  Here                                     
  \be
 \delta {\bf A}^{(5)}       
    =  {1\over s_{45}}\bigg[ {c(12345)\delta \hat n(12;3;45)
   \over \hat s_{12}}+ {c(45123)\delta \hat n(45;1;23)\over s_{23}}               
    +{c(13245)\delta \hat n(13;2;45)\over \hat s_{13}} \bigg]. %(4.38)
\ee
In a familiar way by now, we find the shifts to be
\be
\delta \hat n(12;3;45)=-s_{24}\hat s_{12} \hat f_4, \ \
  \delta \hat n(13;2;45)=s_{24}\hat s_{13} \hat f_4, \ \                                 
 \delta \hat n(45;1;23)=-s_{24} s_{23} \hat f_4.%(4.39)
\ee
Thereupon
 \be
 \delta  {\bf A}^{(5)}=-{s_{24}\over s_{45}}
 \big( c(12345)-c(45123)-c(13245)\big) \hat f_4, %(4.40)
\ee
 which is zero upon manipulating the color Jacobi identities
 a bit.  Similarly, one has   $\delta  {\bf A}^{(5)}_{gr}=0.$  
 
 What we would like to note in this exercise is that the four different
 gauge functions $f_i$ have their respective roles in different
 channels to activate non-trivial gauge transformations.  What is more remarkable is that Bern et al found that the arbitrariness in constructing the five particle $n$'s from the three and four particle ones
is a gauge transformation, which has the form of eq. (4.39).
They gave a definite expression for $-s_{24}\hat f_4$, but it depends on the
$n_i$'s chosen by them.      

There is a particular choice of $f_i$ which will exhaust the
gauge freedom for the shifted numerators.  This is
\be
f_1=-{n(13;4;25)\over s_{13}s_{24}}, \ \
    f_2=-{n(12;4;35)\over s_{24}s_{35}}, \ \
    f_3=-{n(14;2;35)\over s_{24}s_{35}}, \ \
    f_4=-{n(13;2;45)\over s_{13}s_{24}},  %(4.41)
\ee
which gives
\be
\langle N'^{(5)}|=\langle n'_1, \ \ n'_2, \ \ 0, \ \  0, \ \ 0, \ \ 0|,
%(4.42)
\ee
where
\bea
&& n'_1=n(12;3;45)-n(13;425){s_{12}s_{45}\over s_{13}s_{24}}
    +n(12;4;35){s_{45}(s_{12}+s_{24})\over s_{24}s_{35}}\nn\\ &&
     \ \ \ \ \ \ -n(14;2;35){s_{12}s_{45}\over s_{24}s_{35}}
     +n(13;2;45){s_{12}(s_{24}+s_{45})\over  s_{13}s_{24}}, \nn\\ &&
n'_2=n(14;3;25)+n(13;425){s_{14}(s_{24}+s_{25})\over s_{13}s_{24}}
    -n(12;4;35){s_{14}s_{25}\over s_{24}s_{35}}\nn\\ &&
     \ \ \ \ \ \ +n(14;2;35){s_{25}(s_{14}+s_{24})\over s_{24}s_{35}}
     -n(13;2;45){s_{14}s_{25}\over  s_{13}s_{24}}.%(4.43)
\eea
Clearly, the equation      
\be
|A^{(5)}\rangle =  M^{(5)} |N'^{(5)} \rangle %(4.44)
\ee
tells us that we can solve for $n'_1$ and $n'_2$ in terms of any
two of the six color-ordered amplitudes.  For example, these may
be
\bea
&&n'_1=s_{12}\big( s_{25}A(13425)-(s_{15}+s_{25})A(12435)\big),\nn\\ &&
n'_2=s_{25}\big( -(s_{12}+s_{15})A(13425)+s_{12}A(12435)\big).%(4.45)
\eea
When we put these back into the right hand side of eq. (4.44), we
obtain relations amongst the color ordered amplitudes, and
into the equation
\be
{\bf A}^{(5)}_{gr}=\langle \tilde N'^5|M^5|N'^5\rangle %(4.46)
\ee
we have graviton amplitudes in terms of color-ordered gauge
amplitudes, some KLT relations.

%%%%%%%%%%%%%%%%%%%%%%%%%%%%%%%%%%%%%%%%%%%%%%%%%%%%%%%%%%%%%%%%%%%%%%%%%%%%%%%%%%%%%%%%%%%%%%%%%%%%%%%%%%%%%%%%%%%%%%%%%%%%%%%%%%%%%%%%%%    
    \section{Concluding Remarks}
%%%%%%%%%%%%%%%%%%%%%%%%%%%%%%%%%%%%%%%%%%%%%%%%%%%%%%%%%%%%%%%%%%%%%%%%%%%%%%%%%%%%%%%%%%%%%%%%%%%%%%%%%%%%%%%%%%%%%%%%%%%%%%%%%%%%%%%%%     
    
    We have identified in the context of S-matrix theory the
origin of the constraints and the gauge freedom on gluon and
graviton amplitudes.  It is that the propagator matrix $M$
in $|A\rangle=M|N\rangle $ has null eigenvectors $|\lambda ^0_j\rangle.$
Their existence on the one hand reduces the number of
independent entries (color-ordered amplitudes) in $|A\rangle$
and on the other allows one to shift the numerator vector
$|N\rangle \to |N\rangle+\sum_j f_j |\lambda ^0_j\rangle$, a set of gauge
transformations.  Because of the
squaring hypothesis, this freedom transcends to the
graviton amplitudes in the form of KLT relations.

What is even more compelling to investigate this
subject is that in the recursive construction of
scattering amplitudes, there is the necessity and the freedom
to perform a gauge transformation to go from $n$ to $n+1$ particles.
While globally the effects are nil on the fully color dressed gluon
amplitudes and graviton amplitudes, which is meant by gauge
invariance, it has profound implication on
 the semi-local characterization of  interactions, particularly
that aspect which has to do with the duality symmetry between
color kinematics and numerator dynamics.  We have in mind
the promotion of the gauge functions $f_j$ into operator
functionals, or effective Lagrangian.  For example, for the
five gluon scattering, eqs.(4.34), (4.40) and others inform us
that the effective Lagrangian must yield for each channel $s_{ij}$
\bea
\delta ^4 (\sum _a p_a)\delta {\bf A}^{(5)}  &
=& \delta ^4 (\sum _a p_a){1\over s_{ij}}(c(ijklm)+c(ijlmk)
+c(ijmkl))f_{ijklm} \nn\\ &
\sim & \langle 0| \int d^4 x L_{eff}^{(5)}(x)|five\  gluon\  state\rangle,
\eea
where we have replaced $-s_{24}\hat f_j$ by
$f_{ijklm}$ with a more elaborate set of indices to
amplify the color and other attributes.  From dimensional
counting and the factorization of each channel into
$2\to 3$, it is not hard to convert the above up to a constant
into the $L_{eff}^{(5)}(x)$ of Bern et al \cite{Bern:2010ue}.  
This approach takes on additional
significance if through it one can unravel the meaning of the conventional highly non-local non-polynomial gravitational Lagrangians
in perturbative studies.  Hopefully a complementary principle may
ensue.  We shall return to these issues.
\\

{\bf Note Added}: As we prepared our paper for publication, we saw
\cite{BjerrumBohr:2010yc} which has overlap with our work.
\\

{\bf Acknowledgements}

The work of DV is supported in part by a DOE HEP grant DE-FG02-97ER41027
and by a Jeffress Research grant GF12334.

%%%%%%%%%%%%%%%%%%%%%%%%%%%%%%%%%%%%%%%%%%%%%%%%%%%%%%%%%%%%%%%%%%%%%%%%%%%%%%%%%%%%%%%%%%%%%%%%%%%%%%%%%%%%%%%%%%%%%%%%%%%%%%%%%%%%%%%%%%%%%%%%%%%%%%%%%%%%%%
\begin{appendix}
\section{Six Gluon Amplitudes}

There are 24 independent color-ordered gluon amplitudes in the Kleiss-Kuijf basis, which we denote by $A(1 i_2 i_3 i_4 i_5 6)$ with $(i_2,i_3,i_4,i_5)$ equal to a permutation of indices (2,3,4,5). Correspondingly, there are 24 independent numerators which we denote by $n(1j;k;l;m6)$, where as in sections 2 and 3, the external gluons which share a vertex are denoted by numerals which are not separated by a semi-column. The ordering of the gluons is clock-wise. We use the shorthand notation
\bea
&&n(12;3;4;56)=n_1,\qquad n(13; 2;4;56)=n_2,\qquad n(13;4;2;56)=n_3,\qquad
n(13;4;5;26)=n_4\nn\\
&&n(12;4;3;56)=n_5,\qquad n(14;2;3;56)=n_6,\qquad n(14;3;2;56)=n_7,\qquad
n(14;3;5;26)=n_8\nn\\
&&n(12;5;4;36)=n_9,\qquad n(15;2;4;36)=n_{10},\qquad n(15;4;2;36)=n_{11},\qquad
n(15;4;3;26)=n_{12}\nn\\
&&n(12;3;5;46)=n_{13},\qquad n(13;2;5;46)=n_{14},\qquad n(13;5;2;46)=n_{15},\qquad n(13;5;4;26)=n_{16}
\nn\\
&&n(12;4;5;36)=n_{17},\qquad n(14;2;5;36)=n_{18},\qquad n(14;5;2;36)=n_{19},\qquad n(14;5;3;26)=n_{20}\nn\\
&&n(12;5;3;46)=n_{21},\qquad n(15;2;3;46)=n_{22},\qquad n(15;3;2;46)=n_{23},\qquad n(15;3;4;26)=n_{24}.\nn\\
\label{KK6}
\eea
Each of the 24 amplitudes can be written as a sum over 14-type Feynman diagrams. For example, for $A(123456)$ we have
\bea
A(123456)&=&\frac{n_1}{s_{12}s_{123}s_{1234}}-
\frac{n(12;3;6;45)}{s_{12}s_{123}s_{1236}}-
\frac{n(12;6;3;45)}{s_{12}s_{126}s_{1263}}\nn\\
&+&\frac{n(61;2;3;45)}{s_{16}s_{162}s_{1623}}+
\frac{n(12;6;5;34)}{s_{12}s_{126}s_{1265}}+
\frac{n(23;4;5;61)}{s_{23}s_{234}s_{2345}}\nn\\
&-&\frac{n(23;4;1;56)}{s_{23}s_{234}s_{2341}}-
\frac{n(23;1;4;56)}{s_{23}s_{231}s_{2314}}
+\frac{n(34;2;1;56)}{s_{34}s_{342}s_{3421}}\nn\\
&-&\frac{n(34;5;2;61)}{s_{34}s_{345}s_{3452}}
-\frac{n(34;2;5;61)}{s_{34}s_{342}s_{3425}}
+\frac{n(23;1;6;45)}{s_{23}s_{231}s_{2316}}\nn\\
&+&\frac{n(12;34;56)}{s_{12}s_{34}s_{56}}
+\frac{n(61;23;45)}{s_{61}s_{23}s_{45}},
\eea
where the last two terms correspond to snow-flake Feynman diagrams, as their pole structure indicates.
Using that the numerators obey Jacobi-type identities:
\bea
&&n(ij;k;l;mn)=-n(ji;k;l;mn)=-n(ij;k;l;nm), \nn\\
&&n(ij;k;l;mn)+n(ij;k;m;nl)+n(ij;k;n;lm)=0,  \nn\\
&&n(ij;kl;mn)=n(ij;k;l;mn)-n(ij;l;k;mn),
\eea
together with the time-reversed identity
\be
n(ij;k;l;mn)=n(nm;l;k;ji),
\ee
and using that an external line joined with two internal lines at a vertex may be flipped under the internal lines at the price of an additional factor of $(-1)$, we can express the numerators in the color-ordered amplitude $A(123456)$ in the chosen
basis of independent numerators as follows:
\bea
&&n(12;3;6;45)=-n_{1}+n_{13}\nn\\
&&n(12;6;3;45)=-n_1-n_{9}+n_{13}+n_{17}\nn\\
&&n(61;2;3;45)=n_1-n_4+n_9-n_{12}-n_{13}+n_{16}-n_{17}+n_{20}\nn\\
&&n(12;6;5;34)=n_1-n_5+n_{9}-n_{21}\nn\\
&&n(23;4;5;61)=n_1-n_2-n_6+n_7+n_{11}-n_{12}-n_{22}+n_{23}\nn\\
&&n(23;4;1;56)=-n_1+n_2+n_6-n_7\nn\\
&&n(23;1;4;56)=-n_1+n_2\nn\\
&&n(34;2;1;56)=n_1-n_3-n_5+n_7\nn\\
&&n(34;5;2;61)=-n_1+n_4+n_5-n_8-n_9+n_{12}+n_{21}-n_{24}\nn\\
&&n(34;2;5;61)=-n_1+n_3+n_5-n_7-n_{10}+n_{12}+n_{22}-n_{24}\nn\\
&&n(23;1;6;45)=n_1-n_2-n_{13}+n_{14}\nn\\
&&n(12;34;56)=n_1-n_5\nn\\
&&n(61;23;45)=n_1-n_2+n_{11}-n_{12}-n_{13}+n_{14}-n_{19}+n_{20}.
\eea
Similar expressions may be found for the other elements of the Kleiss-Kuijf basis by permutation of indices, and the full propagator matrix $M^{(6)}$ can be constructed in this way. It can be verified that the constraint matrix ${\cal C}^{(6)}$ has six null eigenvectors (while the expressions of these null eigenvectors may be found analytically, they are not particularly illuminating, and we do not give them here), and so, only 18 of the constraints are independent. The propagator matrix $M^{(6)}$ has rank 6 (and thus equal to $(n-3)!$). We have 
separately computed the 18 null eigenvectors of the propagator matrix. Based on our previous discussion of the 4-point and 5-point amplitudes, these null eigenvectors can be chosen such that each 24-component vector has nontrivial entries among the 1,2,3,5,6, 7 set in the chosen Kleiss-Kuijf basis (\ref{KK6}), plus one (and only one) more non-trivial component. For example, 
\bea
&|&-(s_{13}+s_{23})(s_{36}+s_{34}+s_{45}+s_{46}),s_{13}(s_{36}+s_{34}+s_{45}+s_{46}),s_{13}(-s_{14}+s_{36}),0,\nn\\&&(s_{14}+s_{24})s_{35},s_{14}s_{35},(s_{23}+s_{35})s_{14},0,s_{125}s_{36},0,\dots,0\rangle\nn\\
&|&s_{12},s_{23}+s_{12},s_{12}+s_{23}+s_{24},-s_{26},0,\dots,0\rangle
\eea
are such null eigenvectors. For a concrete 6-point MHV amplitude, similar to the 5-point amplitude discussed by Bern et al. \cite{Bern:2010yg}, we can check that the shifts induced by the the null eigenvectors are those needed to reconstruct the 6-point amplitude BCJ numerators (namely those in (\ref{KK6})) via BCFW shifts. 

The fact that the null eigenvectors induce generalized gauge transformations at each channel is a consequence of $\sum_j \hat s_{\alpha} \hat M_{ij} (z) \delta \hat n_j(z)\bigg|_{z=z_\alpha}=0$, for each channel $\alpha$. This is true for any $n$-point amplitude. In principle, the arbitrary functions $f_j$ which multiply the null eigenvector shifts can be appropriately chosen for each channel.

Given the pattern that the numerator matrix exhibits (sums over products of propagators of common $n-3$-tuples formed from adjacent gluon lines between two amplitudes in the Kleiss-Kuijf amplitude basis, with overall sign factors for certain entries: e.g. $M_{12}^{(5)}=1/(s_{15}s_{152}) + 1/(s_{15}s_{154})$ for the numerator entry corresponding to '1'=$A(12345)$ and '2'=$A(14325)$ etc) one would hope that it is possible to write the structure of the null eigenvectors for a generic $n$-point numerator matrix. We plan to return to this question.

\end{appendix}

%%%%%%%%%%%%%%%%%%%%%%%%%%%%%%%%%%%%%%%%%%%%%%%%%%%%%%%%%%%%%%%%%%%%%%%%%%%%%%%%%%%%%%%%%%%%%%%%%%%%%%%%%%%%%%%%%%%%%%%%%%%%%%%%%%%%%%%%%%%%%%%%%%%%%%%%%%%%%

\end{document}